\documentclass[twocolumn,showpacs,aps,prc,superscriptaddress,floatfix]{revtex4}

\usepackage[dvips]{graphicx}

\begin{document}

\title{
Measurement of the Generalized Polarizabilities of
the Proton in Virtual Compton Scattering at 
$Q^2$=0.92 and 1.76 GeV$^2$ :
I. Low Energy Expansion Analysis }

\author{S.~Jaminion}
\affiliation{Universit\'{e} Blaise Pascal/IN2P3, F-63177 Aubi\`{e}re, France}
\author{N.~Degrande}
\affiliation{University of Gent, B-9000 Gent, Belgium}
\author{G.~Laveissi\`{e}re}
\affiliation{Universit\'{e} Blaise Pascal/IN2P3, F-63177 Aubi\`{e}re, France}
\author{C.~Jutier}
\affiliation{Universit\'{e} Blaise Pascal/IN2P3, F-63177 Aubi\`{e}re, France}
\affiliation{Old Dominion University, Norfolk, VA 23529}
\author{L.~Todor}
\affiliation{Old Dominion University, Norfolk, VA 23529}
\author{R.~Di Salvo}
\affiliation{Universit\'{e} Blaise Pascal/IN2P3, F-63177 Aubi\`{e}re, France}
\author{L.~Van Hoorebeke}
\affiliation{University of Gent, B-9000 Gent, Belgium}
\author{L.C.~Alexa}
\affiliation{University of Regina, Regina, SK S4S OA2, Canada}
\author{B.D.~Anderson}
\affiliation{Kent State University, Kent OH 44242}
\author{K.A.~Aniol}
\affiliation{California State University, Los Angeles, CA 90032}
% ...
\collaboration{The Jefferson Lab Hall A Collaboration}
\noaffiliation


\begin{abstract}
Virtual Compton Scattering is studied at
the Thomas Jefferson National Accelerator Facility
at low Center-of-Mass energies, below pion threshold.
Following the Low Energy Theorem for the \ 
$ep \to ep \gamma$ \ process,
we obtain values for the two structure functions 
$P_{LL}-P_{TT}/\epsilon$ and $P_{LT}$ at four-momentum 
transfer squared Q$^2$= 0.92 and 1.76 GeV$^2$.
\end{abstract}



\maketitle

%
%


The electric and magnetic polarizabilities of the nucleon
reflect its response to a static electromagnetic field.  These are 
fundamental observables of the ground state, closely related to
the entire excitation spectrum of the nucleon.  The polarizabilities
of the proton have been measured in Real Compton Scattering 
(RCS) experiments $\gamma p \rightarrow \gamma p$; 
see e.g.~\cite{OlmosdeLeon:2001zn}.
Contrary to atomic polarizabilities, which are of the size
of the atomic volume~\cite{Dzuba:1997df},  
the proton electric polarizability is
much smaller than one cubic fm, the volume scale of a nucleon.
In a simplified harmonic oscillator model,
such a small electric polarizability is a natural indication of the
intrinsic relativistic character of the nucleon. 
The smallness of the proton magnetic polarizability $\beta_M$
relative to $\alpha_E$ reflects a strong cancellation of 
para- and dia-magnetism in the proton.      


A theoretical study of the Virtual Compton Scattering (VCS)
reaction  $\gamma^* p \rightarrow \gamma p$, at threshold
but at arbitrary virtuality $Q^2$ of the initial photon,
led to the concept of Generalized Polarizabilities 
(GP)~\cite{Guichon:1995pu}.
The GPs measure the spatial variation of the polarization of the
proton, induced by external electric- or magnetic-dipole fields
as a function of $Q^2$.
They provide an original way to probe the nucleon structure.
After the NE-18 experiment~\cite{vandenBrand:1995wj} and the
pioneering VCS experiment at MAMI~\cite{Roche:2000ng},
the E93-050 experiment~\cite{Bertin:1993} was performed at higher
energies at Jefferson Lab (JLab).
The studied reaction channel is the
exclusive photon electroproduction  $ep \to ep \gamma$. 
In this Letter, we report
the results of the low-energy expansion (LEX) analysis 
of this experiment, based on the deviations of the cross
section with respect to the Low Energy Theorem. 
In a companion Letter
(referred to as II) we report the results of a Dispersion Relation
analysis of our data~\cite{Laveissiere:2003dr}.

According to  P. Guichon 
\textit{et al.}~\cite{Guichon:1995pu},
the unpolarized cross section for the reaction  
$e p \rightarrow e p \gamma$ at small momentum $q'$ of the final
photon can be written:
\begin{eqnarray}
 d^5 \sigma ^{EXP} \ = \  
 d^5 \sigma ^{BH+Born}  \ + \ q' \phi \Psi_0 \ + \ {\cal O } (q'^2) ,
 \nonumber \\ 
\Psi_0 \ =  \ v_1 
(P_{LL} - {\displaystyle 1 \over \displaystyle \epsilon} P_{TT}) 
\ + \ v_2 P_{LT}  
\label{eq01}
\end{eqnarray} 
where $\phi, v_1, v_2$ are kinematical coefficients,
$q'$ is the modulus of the three-momentum 
of the final photon in the $(\gamma p)$ CM frame
and $\epsilon$ is the virtual photon polarization.
The differential elements of $d^5 \sigma$ are the
scattered electron momentum and solid angle in the lab
frame and the proton solid angle in the $(\gamma p)$ CM frame.
$\Psi_0$ contains the effect of the GPs through
the structure functions $P_{LL}$, $P_{TT}$, $P_{LT}$.
$\Psi_0$ represents
the effect of GPs to first order in $q'$, or equivalently
the truncation to electric- or magnetic-dipole radiation. 
It has been shown to contain six independent (dipole) 
GPs~\cite{Drechsel:1997ag,Drechsel:1998xv}, combined into the
structure functions.
$d^5 \sigma ^{BH+Born}$
corresponds to the coherent sum of the Bethe-Heitler (BH)
and the VCS Born amplitudes. It depends only on the
elastic form factors of the proton and is a particular case
of Low's low-energy theorem~\cite{Low:1958sn} 
for threshold photon production. 

The goal of the measurement, 
performed at fixed $\epsilon$,
is to extract the two structure functions 
$P_{LL} -P_{TT}/\epsilon$ and $P_{LT}$ by
extrapolating the quantity
\begin{eqnarray}
 (d^5 \sigma ^{EXP} -  d^5 \sigma ^{BH+Born}) / ( q' \phi) 
\label{eq01bis}
\end{eqnarray} 
to $q'=0$. This is done at a fixed value of $q$, the
CM three-momentum of the VCS virtual photon.
Thus, for the small but non-vanishing values of $q'$ 
that we consider, the corresponding squared momentum 
$Q^2=q^2-q_0^2$ depends on $q'$ (through 
the CM virtual photon energy $q_0$).
Though this dependence is very weak for our kinematical 
conditions, we point out that in the following,
$Q^2$ will refer to the value corresponding to 
$q'=0$, that is
$Q^2= 2 M_p \cdot ( \sqrt{M_p^2 + q^2} -M_p)$.
Without listing the complete expression of the structure functions 
in terms of the GPs~\cite{Guichon:1998xv,Roche:2000ng}, 
we just mention that $P_{LL}$ is
proportional to the electric GP $\alpha_E (Q^2) $ and that
$P_{LT}$ contains the magnetic GP $\beta_M(Q^2)$.
When $Q^2$ tends to zero, these two GPs give the
well-known electric and magnetic polarizabilities
obtained in RCS. The structure function $P_{TT}$  is a combination
of spin-flip GPs.


The apparatus and running conditions of the JLab experiment
are detailed 
elsewhere~\cite{Laveissiere:2003pi,Degrande:2001th,
Jaminion:2001th,Jutier:2001th,Laveissiere:2001th,Todor:2000th}. 
Here a brief summary is given. 
An electron beam of 4.030 GeV energy 
was directed onto a 15 cm liquid hydrogen target.
The two Hall A Spectrometers were used to 
detect the scattered electron and the outgoing proton in coincidence,
allowing the identification of the exclusive 
reaction $ ep \to ep \gamma$ \ by  the 
missing-mass technique.
This experiment makes use of the full capabilities of the 
accelerator and the Hall A instrumentation~\cite{Alcorn:2003}:
100\% duty cycle, high resolution spectrometers, 
high luminosities (up to $4 \times 10^{38}$ cm$^{-2} \cdot s^{-1}$).


The data for the LEX analysis are divided into two subsets 
corresponding to data taking in two different ranges of
$Q^2$: [0.85, 1.15] and [1.6, 2.1] GeV$^2$. 
Spectrometer optics are optimized for the experiment,
and dedicated  procedures allow adjustment of
the main experimental offsets in energies, angles 
and positions.
For each event, variables such as
$q'$, or the CM polar and azimuthal angles $ \theta$ and
$\varphi$ of the outgoing photon w.r.t. $\vec q$, 
are obtained by reconstructing the missing particle. 
The acceptance calculation is provided by 
a dedicated Monte-Carlo simulation~\cite{VanHoorebeke} 
including resolution effects. 
Radiative corrections are applied following the 
exponentiation method of Ref.~\cite{Vanderhaeghen:2000ws}, the
acceptance-dependent part being implemented in the simulation.
The acceptance calculation requires
a realistic shape for the sampling cross section. 
For this reason, simulated events are generated using
the (BH+Born) cross section plus a first-order GP effect,
introduced iteratively.
Clean event samples are obtained by applying the event selection
method. It involves
simple cuts, e.g. a window around 0 in missing mass 
squared to select the photon peak, or the condition
$q'<126$ MeV/c to stay below pion threshold. Other cuts are less
trivial; special attention was paid to
obtain a well-defined acceptance and to
eliminate protons punching through the spectrometer
entrance collimator.
The same cuts are applied to experimental and simulated events.
The absolute normalization of the cross section is obtained
from the knowledge of the beam charge, target density, 
and detectors efficiencies.
More details can be found in 
Refs.~\cite{Degrande:2001th,
Jaminion:2001th,Jutier:2001th}.

\begin{figure}[t]
\includegraphics[width=8.6cm,height=5.7cm]{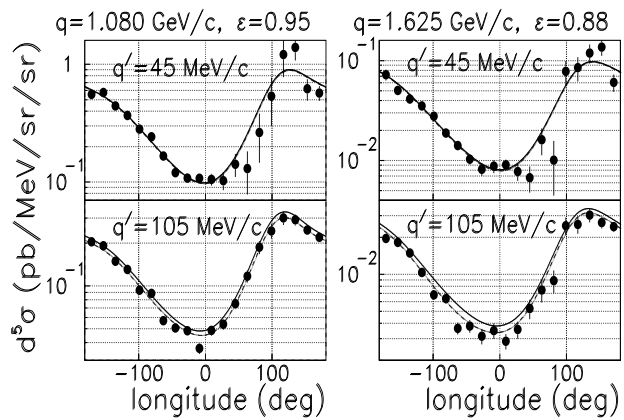}
\caption{\label{seceff}
$(ep \to ep \gamma)$ cross section
for the lowest and highest $q'$ bin, at  
40$^{\circ}$ out-of-plane 
(including symmetry w.r.t. the lepton plane). 
Only statistical errors are shown.
Here the out-of-plane angle (or latitude) is the polar angle
of the outgoing photon when the polar axis 
is chosen perpendicular to the lepton plane.
The abscissa is the azimuthal angle (or longitude)
using this convention; negative values are defined as in 
Ref.~\cite{Roche:2000ng}. The full curve
is the (BH+Born) cross section, the dashed curve includes
the first-order GP effect (Eq.~\ref{eq01}) as fitted in this
analysis.
}
\end{figure}

For each of the two data sets, the
photon electroproduction cross section is determined
at a fixed value of $q$ (1.080 and 1.625 GeV/c)
and $\epsilon$ (0.95 and 0.88).
Events are binned in $q'$ (30 MeV/c wide), 
and in $\theta$ and $\varphi$ (see Fig.~\ref{seceff}).
The advantage of the JLab experiment is that it
produces a large out-of-plane acceptance, thanks to
the Lorentz boost (CM $\to$ lab) focusing 
the outgoing  proton around $\vec q$.  
An example of out-of-plane cross-section data
is given in Fig.~\ref{seceff}.
At the lowest $q'$ of 45 MeV/c, the GP effect is very small
($<$ 3\% of $d^5 \sigma$) and not visible on the plot; 
at the highest $q'$ below pion threshold (105 MeV/c) 
it reaches 10-15\% . 
We confirmed the central value of the absolute 
normalization of the experiment 
to the 1\% level in the lowest $q'$-bin,
where the experimental cross section is close to the
known (BH+Born) cross section.
For this test and throughout our analyses, the
(BH+Born) cross section is calculated using
the parametrization of Ref.~\cite{Brash:2001qq}
for the proton EM form factors.

\begin{figure}[b]
\includegraphics[width=8.6cm,height=5.7cm]{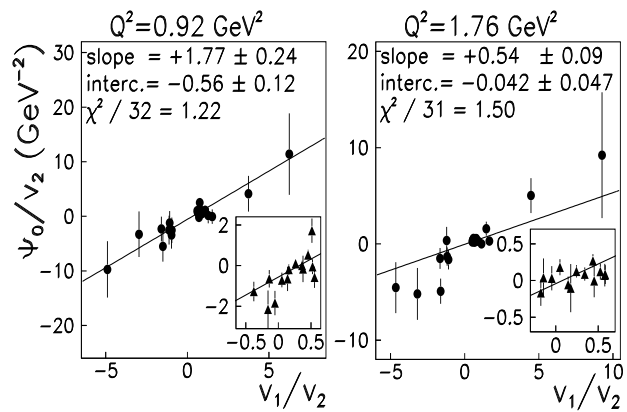}
\caption{\label{fitgps}
A graphical representation of the LEX fit (straight line) 
for each data set. 
Black circles correspond to out-of-plane data,
and the inner plot is a zoom on the lepton plane
data (triangles).  $\Psi_0, v_1$ and $v_2$ are defined in the text. }
\end{figure}

The method to extract the structure functions 
is deduced from Eq.~\ref{eq01}.
For each bin in $(\theta, \varphi)$, one measures 
$d^5 \sigma ^{EXP}$ at several finite values of $q'$,
and extrapolates the quantity 
$\Delta {\cal M} = 
 (d^5 \sigma ^{EXP} -  d^5 \sigma ^{BH+Born})/( \phi q' )$
to $q'=0$, yielding the value of $\Psi_0$. 
In our data, $\Delta {\cal M}$ does not exhibit 
any significant $q'$-dependence, so the extrapolation to $q'=0$
is done in each bin in $(\theta, \varphi)$ 
by averaging $\Delta {\cal M}$ over $q'$.
The resulting $\Psi_0$ term is then fitted as a linear combination 
of two free parameters, which are 
the structure functions $P_{LL}-P_{TT}/\epsilon$ and 
$P_{LT}$ (see Fig.~\ref{fitgps}).
Their  statistical error 
is given by the fit, and their systematic error 
is calculated from four sources added quadratically:
1) $\pm$ 2 MeV on beam energy, 
2) $\pm$ 0.5 mrad on horizontal angles, 
3) $\pm$ 2.3\% on overall absolute cross section normalization and 
4) $\pm$ 2\% due to possible cross section shape distortions.
The value of the reduced $\chi^2$ of the fit
(1.22 and 1.50, cf. Fig.~\ref{fitgps}) is one way to express 
that the LEX holds reasonably well at our kinematics.
 
Our results, presented 
in Table~\ref{lexresults} together with the previous 
results at lower $Q^2$, show a
strong fall-off of the structure functions with 
momentum transfer, similarly to form factors. 
\begin{table}[t]
\caption{\label{lexresults} Compilation of 
the VCS structure functions. The first line corresponds
to the RCS result~\cite{OlmosdeLeon:2001zn}.
In all cases the first error is statistical, 
and the second one is the total systematic error.
 }
\begin{ruledtabular}
\begin{tabular}{ccccc}
Ref. & $Q^2$  & $\epsilon$  &
$P_{LL}-P_{TT}/\epsilon$ & $P_{LT}$ \\
\ & (GeV$^2$) & \ & (GeV$^{-2}$) & (GeV$^{-2}$) \\
 \hline
\multicolumn{5}{c}{Previous experiments} \\
\hline
\cite{OlmosdeLeon:2001zn} & 0 &  \ & 81.3  & -5.4   \\
\ & \  & \  
& { $\pm$ 2.0 $\pm$ 3.4 }
& { \small $\pm$  1.3 $\pm$ 1.9 } \\
\cite{Roche:2000ng} & 0.33 & 0.62  & 23.7 & -5.0 \\
\ & \ & \ 
& { $\pm$ 2.2 $\pm$ 4.3 }
& { $\pm$ 0.8 $\pm$ 1.8 } \\
\hline
\multicolumn{5}{c}{ This experiment}  \\ 
\hline
\  & 0.92 & 0.95  & \textbf{1.77} & \textbf{-0.56} \\
\ & \ & \  
& { $\pm$ 0.24 $\pm$ 0.70 }
& { $\pm$ 0.12 $\pm$ 0.17 } \\
\ & 1.76 & 0.88  & \textbf{0.54} & \textbf{-0.042} \\
\ & \ & \ 
& { $\pm$ 0.09 $\pm$ 0.20 }
& { $\pm$ 0.047 $\pm$ 0.055 } \\
\end{tabular}
\end{ruledtabular}
\end{table}
This behavior is expected
since GPs can be seen as ``form factors of
a nucleon embedded in an EM field''.
From a comparison with theoretical predictions, one should point out
that most model calculations of the GPs, e.g. 
Refs.~\cite{Hemmert:1999pz,Kao:2002cn,Pasquini:2000ue}
are valid at much smaller $Q^2$ than 
the ones covered in this experiment.
One notable exception is the
Dispersion Relation (DR) model~\cite{Drechsel:2002ar};
our extraction of VCS structure functions 
in the same experiment,
based on the DR approach, 
is the subject of a companion Letter~\cite{Laveissiere:2003dr}.
The results presented here should stimulate 
theoretical calculations of GPs at intermediate and high $Q^2$.
It would be of great interest
if the models that predict the form factors $G_E$ and $G_M$ 
at $Q^2 \ge$ 1 GeV$^2$ could give predictions of the GPs
in the same $Q^2$ range. 


In summary, we have studied the process $e p \to e p \gamma$
using the unique capabilities of the JLab electron beam and the 
Hall A instrumentation. We have measured the two
unpolarized structure functions of Virtual Compton Scattering
$P_{LL}-P_{TT}/\epsilon$ and $P_{LT}$, providing a new piece
of information on the 
electromagnetic structure of the nucleon, namely on the
 behavior of the proton Generalized Polarizabilities
in the range $Q^2$ $\sim$ 1-2 GeV$^2$.
Other VCS experiments at low energy will measure these 
structure functions
(MIT-Bates~\cite{Shaw:1997}, MAMI~\cite{Merkel:2000}) and 
double polarization is foreseen~\cite{Merkel:2000} 
as a way to separate the six lowest order GPs.

We thank the JLab accelerator staff and the Hall A technical staff
for their dedication.
This work was supported by DOE contract DE-AC05-84ER40150 under
which the Southeastern Universities Research Association (SURA)
operates the Thomas Jefferson National Accelerator Facility. We
acknowledge additional grants from the US DOE and NSF, the French
Centre National de la Recherche Scientifique and Commissariat \`a
l'Energie Atomique, the Conseil R\'egional d'Auvergne, the
FWO-Flanders (Belgium) and the BOF-Gent University. 
We thank M.Vanderhaeghen and B.Pasquini for fruitful discussions
and the organization of VCS workshops at ECT* (Trento).

\bibliography{/users/divers/cebaf/vcs/publications/struc_func/common}

\begin{thebibliography}{27}
\expandafter\ifx\csname natexlab\endcsname\relax\def\natexlab#1{#1}\fi
\expandafter\ifx\csname bibnamefont\endcsname\relax
  \def\bibnamefont#1{#1}\fi
\expandafter\ifx\csname bibfnamefont\endcsname\relax
  \def\bibfnamefont#1{#1}\fi
\expandafter\ifx\csname citenamefont\endcsname\relax
  \def\citenamefont#1{#1}\fi
\expandafter\ifx\csname url\endcsname\relax
  \def\url#1{\texttt{#1}}\fi
\expandafter\ifx\csname urlprefix\endcsname\relax\def\urlprefix{URL }\fi
\providecommand{\bibinfo}[2]{#2}
\providecommand{\eprint}[2][]{\url{#2}}

\bibitem[{\citenamefont{Olmos~de Leon et~al.}(2001)}]{OlmosdeLeon:2001zn}
\bibinfo{author}{\bibfnamefont{V.}~\bibnamefont{Olmos~de Leon}}
  \bibnamefont{et~al.}, \bibinfo{journal}{Eur. Phys. J.}
  \textbf{\bibinfo{volume}{A10}}, \bibinfo{pages}{207} (\bibinfo{year}{2001}).

\bibitem[{\citenamefont{Dzuba et~al.}(1997)\citenamefont{Dzuba, Flambaum, and
  Sushkov}}]{Dzuba:1997df}
\bibinfo{author}{\bibfnamefont{V.~A.} \bibnamefont{Dzuba}},
  \bibinfo{author}{\bibfnamefont{V.~V.} \bibnamefont{Flambaum}},
  \bibnamefont{and} \bibinfo{author}{\bibfnamefont{O.~P.}
  \bibnamefont{Sushkov}} (\bibinfo{year}{1997}), \eprint{hep-ph/9709251}.

\bibitem[{\citenamefont{Guichon et~al.}(1995)\citenamefont{Guichon, Liu, and
  Thomas}}]{Guichon:1995pu}
\bibinfo{author}{\bibfnamefont{P.~A.~M.} \bibnamefont{Guichon}},
  \bibinfo{author}{\bibfnamefont{G.~Q.} \bibnamefont{Liu}}, \bibnamefont{and}
  \bibinfo{author}{\bibfnamefont{A.~W.} \bibnamefont{Thomas}},
  \bibinfo{journal}{Nucl. Phys.} \textbf{\bibinfo{volume}{A591}},
  \bibinfo{pages}{606} (\bibinfo{year}{1995}).

\bibitem[{\citenamefont{van~den Brand et~al.}(1995)}]{vandenBrand:1995wj}
\bibinfo{author}{\bibfnamefont{J.~F.~J.} \bibnamefont{van~den Brand}}
  \bibnamefont{et~al.}, \bibinfo{journal}{Phys. Rev.}
  \textbf{\bibinfo{volume}{D52}}, \bibinfo{pages}{4868} (\bibinfo{year}{1995}).

\bibitem[{\citenamefont{Roche et~al.}(2000)}]{Roche:2000ng}
\bibinfo{author}{\bibfnamefont{J.}~\bibnamefont{Roche}} \bibnamefont{et~al.},
  \bibinfo{journal}{Phys. Rev. Lett.} \textbf{\bibinfo{volume}{85}},
  \bibinfo{pages}{708} (\bibinfo{year}{2000}).

\bibitem[{\citenamefont{Bertin et~al.}(1993)\citenamefont{Bertin, Hyde-Wright,
  Guichon et~al.}}]{Bertin:1993}
\bibinfo{author}{\bibfnamefont{P.~Y.} \bibnamefont{Bertin}},
  \bibinfo{author}{\bibfnamefont{C.}~\bibnamefont{Hyde-Wright}},
  \bibinfo{author}{\bibfnamefont{P.~A.~M.} \bibnamefont{Guichon}},
  \bibnamefont{et~al.} (\bibinfo{year}{1993}), \bibinfo{note}{experiment
  E93-050},
  \urlprefix\url{http://hallaweb.jlab.org/experiment/E93-050/vcs.html}.

\bibitem[{\citenamefont{Laveissi\`ere et~al.}()}]{Laveissiere:2003dr}
\bibinfo{author}{\bibfnamefont{G.}~\bibnamefont{Laveissi\`ere}}
  \bibnamefont{et~al.}, \bibinfo{note}{{Letter} II, to be submitted to Phys.
  Rev. Lett.}

\bibitem[{\citenamefont{Drechsel et~al.}(1997)\citenamefont{Drechsel,
  Knochlein, Metz, and Scherer}}]{Drechsel:1997ag}
\bibinfo{author}{\bibfnamefont{D.}~\bibnamefont{Drechsel}},
  \bibinfo{author}{\bibfnamefont{G.}~\bibnamefont{Knochlein}},
  \bibinfo{author}{\bibfnamefont{A.}~\bibnamefont{Metz}}, \bibnamefont{and}
  \bibinfo{author}{\bibfnamefont{S.}~\bibnamefont{Scherer}},
  \bibinfo{journal}{Phys. Rev.} \textbf{\bibinfo{volume}{C55}},
  \bibinfo{pages}{424} (\bibinfo{year}{1997}).

\bibitem[{\citenamefont{Drechsel et~al.}(1998)\citenamefont{Drechsel,
  Knochlein, Korchin, Metz, and Scherer}}]{Drechsel:1998xv}
\bibinfo{author}{\bibfnamefont{D.}~\bibnamefont{Drechsel}},
  \bibinfo{author}{\bibfnamefont{G.}~\bibnamefont{Knochlein}},
  \bibinfo{author}{\bibfnamefont{A.~Y.} \bibnamefont{Korchin}},
  \bibinfo{author}{\bibfnamefont{A.}~\bibnamefont{Metz}}, \bibnamefont{and}
  \bibinfo{author}{\bibfnamefont{S.}~\bibnamefont{Scherer}},
  \bibinfo{journal}{Phys. Rev.} \textbf{\bibinfo{volume}{C57}},
  \bibinfo{pages}{941} (\bibinfo{year}{1998}).

\bibitem[{\citenamefont{Low}(1958)}]{Low:1958sn}
\bibinfo{author}{\bibfnamefont{F.~E.} \bibnamefont{Low}},
  \bibinfo{journal}{Phys. Rev.} \textbf{\bibinfo{volume}{110}},
  \bibinfo{pages}{974} (\bibinfo{year}{1958}).

\bibitem[{\citenamefont{Guichon and Vanderhaeghen}(1998)}]{Guichon:1998xv}
\bibinfo{author}{\bibfnamefont{P.~A.~M.} \bibnamefont{Guichon}}
  \bibnamefont{and}
  \bibinfo{author}{\bibfnamefont{M.}~\bibnamefont{Vanderhaeghen}},
  \bibinfo{journal}{Prog. Part. Nucl. Phys.} \textbf{\bibinfo{volume}{41}},
  \bibinfo{pages}{125} (\bibinfo{year}{1998}).

\bibitem[{\citenamefont{Laveissi\`ere et~al.}(2003)}]{Laveissiere:2003pi}
\bibinfo{author}{\bibfnamefont{G.}~\bibnamefont{Laveissi\`ere}}
  \bibnamefont{et~al.} (\bibinfo{year}{2003}), \bibinfo{note}{to be submitted
  to Phys. Rev. C}, \eprint{nucl-ex/0308009}.

\bibitem[{\citenamefont{Degrande}(2001)}]{Degrande:2001th}
\bibinfo{author}{\bibfnamefont{N.}~\bibnamefont{Degrande}}, Ph.D. thesis,
  \bibinfo{school}{Gent University} (\bibinfo{year}{2001}).

\bibitem[{\citenamefont{Jaminion}(2001)}]{Jaminion:2001th}
\bibinfo{author}{\bibfnamefont{S.}~\bibnamefont{Jaminion}}, Ph.D. thesis,
  \bibinfo{school}{Universit\'e Blaise Pascal} (\bibinfo{year}{2001}),
  \bibinfo{note}{{D}U 1259}.

\bibitem[{\citenamefont{Jutier}(2001)}]{Jutier:2001th}
\bibinfo{author}{\bibfnamefont{C.}~\bibnamefont{Jutier}}, Ph.D. thesis,
  \bibinfo{school}{Old Dominion University and Universit\'e Blaise Pascal}
  (\bibinfo{year}{2001}), \bibinfo{note}{{D}U 1298}.

\bibitem[{\citenamefont{Laveissi\`ere}(2001)}]{Laveissiere:2001th}
\bibinfo{author}{\bibfnamefont{G.}~\bibnamefont{Laveissi\`ere}}, Ph.D. thesis,
  \bibinfo{school}{Universit\'e Blaise Pascal} (\bibinfo{year}{2001}),
  \bibinfo{note}{{D}U 1309}.

\bibitem[{\citenamefont{Todor}(2000)}]{Todor:2000th}
\bibinfo{author}{\bibfnamefont{L.}~\bibnamefont{Todor}}, Ph.D. thesis,
  \bibinfo{school}{Old Dominion University} (\bibinfo{year}{2000}).

\bibitem[{\citenamefont{Alcorn et~al.}()}]{Alcorn:2003}
\bibinfo{author}{\bibfnamefont{J.}~\bibnamefont{Alcorn}} \bibnamefont{et~al.},
  \bibinfo{note}{accepted by NIM A}.

\bibitem[{\citenamefont{Van~Hoorebeke et~al.}()}]{VanHoorebeke}
\bibinfo{author}{\bibfnamefont{L.}~\bibnamefont{Van~Hoorebeke}}
  \bibnamefont{et~al.}, \bibinfo{note}{to be submitted to NIM A}.

\bibitem[{\citenamefont{Vanderhaeghen et~al.}(2000)\citenamefont{Vanderhaeghen,
  Friedrich, Lhuillier, Marchand, Van~Hoorebeke, and Van~de
  Wiele}}]{Vanderhaeghen:2000ws}
\bibinfo{author}{\bibfnamefont{M.}~\bibnamefont{Vanderhaeghen}},
  \bibinfo{author}{\bibfnamefont{J.~M.} \bibnamefont{Friedrich}},
  \bibinfo{author}{\bibfnamefont{D.}~\bibnamefont{Lhuillier}},
  \bibinfo{author}{\bibfnamefont{D.}~\bibnamefont{Marchand}},
  \bibinfo{author}{\bibfnamefont{L.}~\bibnamefont{Van~Hoorebeke}},
  \bibnamefont{and} \bibinfo{author}{\bibfnamefont{J.}~\bibnamefont{Van~de
  Wiele}}, \bibinfo{journal}{Phys. Rev.} \textbf{\bibinfo{volume}{C62}},
  \bibinfo{pages}{025501} (\bibinfo{year}{2000}).

\bibitem[{\citenamefont{Brash et~al.}(2002)\citenamefont{Brash, Kozlov, Li, and
  Huber}}]{Brash:2001qq}
\bibinfo{author}{\bibfnamefont{E.~J.} \bibnamefont{Brash}},
  \bibinfo{author}{\bibfnamefont{A.}~\bibnamefont{Kozlov}},
  \bibinfo{author}{\bibfnamefont{S.}~\bibnamefont{Li}}, \bibnamefont{and}
  \bibinfo{author}{\bibfnamefont{G.~M.} \bibnamefont{Huber}},
  \bibinfo{journal}{Phys. Rev.} \textbf{\bibinfo{volume}{C65}},
  \bibinfo{pages}{051001} (\bibinfo{year}{2002}).

\bibitem[{\citenamefont{Hemmert et~al.}(2000)\citenamefont{Hemmert, Holstein,
  Knochlein, and Drechsel}}]{Hemmert:1999pz}
\bibinfo{author}{\bibfnamefont{T.~R.} \bibnamefont{Hemmert}},
  \bibinfo{author}{\bibfnamefont{B.~R.} \bibnamefont{Holstein}},
  \bibinfo{author}{\bibfnamefont{G.}~\bibnamefont{Knochlein}},
  \bibnamefont{and} \bibinfo{author}{\bibfnamefont{D.}~\bibnamefont{Drechsel}},
  \bibinfo{journal}{Phys. Rev.} \textbf{\bibinfo{volume}{D62}},
  \bibinfo{pages}{014013} (\bibinfo{year}{2000}).

\bibitem[{\citenamefont{Kao and Vanderhaeghen}(2002)}]{Kao:2002cn}
\bibinfo{author}{\bibfnamefont{C.~W.} \bibnamefont{Kao}} \bibnamefont{and}
  \bibinfo{author}{\bibfnamefont{M.}~\bibnamefont{Vanderhaeghen}},
  \bibinfo{journal}{Phys. Rev. Lett.} \textbf{\bibinfo{volume}{89}},
  \bibinfo{pages}{272002} (\bibinfo{year}{2002}).

\bibitem[{\citenamefont{Pasquini et~al.}(2001)\citenamefont{Pasquini, Scherer,
  and Drechsel}}]{Pasquini:2000ue}
\bibinfo{author}{\bibfnamefont{B.}~\bibnamefont{Pasquini}},
  \bibinfo{author}{\bibfnamefont{S.}~\bibnamefont{Scherer}}, \bibnamefont{and}
  \bibinfo{author}{\bibfnamefont{D.}~\bibnamefont{Drechsel}},
  \bibinfo{journal}{Phys. Rev.} \textbf{\bibinfo{volume}{C63}},
  \bibinfo{pages}{025205} (\bibinfo{year}{2001}).

\bibitem[{\citenamefont{Drechsel et~al.}(2003)\citenamefont{Drechsel, Pasquini,
  and Vanderhaeghen}}]{Drechsel:2002ar}
\bibinfo{author}{\bibfnamefont{D.}~\bibnamefont{Drechsel}},
  \bibinfo{author}{\bibfnamefont{B.}~\bibnamefont{Pasquini}}, \bibnamefont{and}
  \bibinfo{author}{\bibfnamefont{M.}~\bibnamefont{Vanderhaeghen}},
  \bibinfo{journal}{Phys. Rept.} \textbf{\bibinfo{volume}{378}},
  \bibinfo{pages}{99} (\bibinfo{year}{2003}).

\bibitem[{\citenamefont{Shaw et~al.}(1997)\citenamefont{Shaw, Miskimen
  et~al.}}]{Shaw:1997}
\bibinfo{author}{\bibfnamefont{J.}~\bibnamefont{Shaw}},
  \bibinfo{author}{\bibfnamefont{R.}~\bibnamefont{Miskimen}},
  \bibnamefont{et~al.} (\bibinfo{year}{1997}), \bibinfo{note}{{MIT-Bates}
  proposal E97-03}.

\bibitem[{\citenamefont{Merkel and d'Hose}(2000)}]{Merkel:2000}
\bibinfo{author}{\bibfnamefont{H.}~\bibnamefont{Merkel}} \bibnamefont{and}
  \bibinfo{author}{\bibfnamefont{N.}~\bibnamefont{d'Hose}}
  (\bibinfo{year}{2000}), \bibinfo{note}{spokspersons, MAMI Proposal}.

\end{thebibliography}

\end{document}